# OCCIA LAB

*To discover the causes of social, economic and technological change*

CocciaLAB Working Paper 2017 – No. 24

# DISRUPTIVE FIRMS

Mario COCCIA .

ARIZONA STATE UNIVERSITY
CocciaLAB is at the Center for Social Dynamics and Complexity
Interdisciplinary Science and Technology Building 1 (ISBT1)
550 E. Orange Street, Tempe- AZ 85287-4804 USA

and

CNR -- NATIONAL RESEARCH COUNCIL OF ITALY
Via Real Collegio, 30-10024, Moncalieri (TO), Italy

*E*-mail: mario.coccia@cnr.it

# DISRUPTIVE FIRMS

*Mario Coccia .[1]*
ARIZONA STATE UNIVERSITY &
CNR -- NATIONAL RESEARCH COUNCIL OF ITALY
*E*-mail: mario.coccia@cnr.it
Mario Coccia ORCID: http://orcid.org/0000-0003-1957-6731

**ABSTRACT**

This study proposes the concept of *disruptive firms*: they are firms with market leadership that deliberate introduce new and improved generations of durable goods that destroy, directly or indirectly, similar products present in markets in order to support their competitive advantage and/or market leadership. These disruptive firms support technological and industrial change and induce consumers to buy new products to adapt to new socioeconomic environment. In particular, disruptive firms generate and spread path-breaking innovations in order to achieve and sustain the goal of a (temporary) profit monopoly. This organizational behaviour and strategy of disruptive firms support technological change. This study can be useful for bringing a new perspective to explain and generalize one of the determinants that generates technological and industrial change. Overall, then this study suggests that one of the general sources of technological change is due to disruptive firms (subjects), rather than disruptive technologies (objects), that generate market shifts in a Schumpeterian world of innovation-based competition.

---

[1] I gratefully acknowledge financial support from the CNR - National Research Council of Italy for my visiting at Arizona State University (Grant CNR - NEH Memorandum Grant n. 0072373-2014 and n. 0003005-2016) where this research started in 2016. The author declares that he has no relevant or material financial interests that relate to the research discussed in this paper.

**Introduction**

Current economies show the advent of many technological advances in information technology, biotechnology, nanotechnology, etc. that generate corporate, industrial and economic change (Arora et al., 2001; Henderson and Clark, 1990; Nicholson et al., 1990; Teece et al., 1997; Van de Ven at al., 2008; von Hippel, 1988). The literature in these research fields has suggested several approaches to explain the technological and industrial change, such as the theory by Christensen (1997, 2006) that introduces the concept of disruptive technologies of new entrants that disrupt the competitive advantage of incumbents in the presence market dynamisms. This theory explains the industrial change with the interplay between incumbent and entrant firms that can generate path-breaking technologies[2]. While the validity of certain of these studies may be debated, it is clear that there are at least some facts about industrial change that theory of disruptive technologies has trouble explaining. As a matter of fact, current dynamics of industries shows that new entrants can generate disruptive technologies but their development and diffusion between markets have more and more economic barriers (Coccia, 2016; 2017). This paper suggests that industrial change is driven by specific subjects -disruptive firms, rather than disruptive technologies *per se*. This study can be useful for bringing a new perspective to explain and generalize one of the sources of technological change that is represented by specific firms that have the potential to generate and/or to develop radical innovations that disrupt current products in markets and support industrial, economic and social change.

In order to position this study in existing approaches, the paper develops the theoretical framework in next section.

---

[2] Cf., Ansari et al., 2016; Baatartogtokh, 2015; Chesbrough and Rosenbloom, 2002; Christensen, 1997, 2006; Christensen *et al.,* 2015; Danneels, 2004, 2006; Gilbert and Bower, 2002; Hill and Rothaermel, 2003; Jenkins, 2010; King and Garud *et al.,* 2015; Ryan and Tipu, 2013; Tellis, 2006; Wessel and Christensen, 2012.

**Theoretical framework**

Many industries are characterized by incumbents that focus mainly on improving their products and services (usually most profitable), and entrants that endeavor to develop new technologies in market segments, delivering market performance that incumbents' mainstream customers require (Christensen et al., 2015; Christensen, 1997). In this context, Christensen (1997) argues that disruptive innovations generate significant shifts in markets (cf., Henderson, 2006). In particular, disruptive innovations are generated by small firms with fewer resources that successfully challenge established incumbent businesses (Christensen et al., 2015). New firms can generate competence-destroying discontinuities that increase the environmental turbulence, whereas incumbents focus mainly on competence-enhancing discontinuities that decrease the turbulence in markets (cf., Tushman and Anderson, 1986). Scholars also argue that the ability of incumbents to develop and to market disruptive innovations is due to their specific ambidexterity: competence-destroying and competence-enhancing based on simultaneous exploratory and exploitative activities to support both incremental and radical innovations (Danneels, 2006; Durisin and Todorova, 2012; Lin and McDonough III, 2014; O'Reilly III and Tushman, 2004, 2008; *cf.,* Henderson, 2006; Madsen and Leiblein, 2015). Disruptive innovations generate main effects both for consumers and producers in markets and society (Markides, 2006, pp. 22-23; Markides and Geroski, 2005). In general, disruptive innovations change habits of consumers in markets and undermine the competences and complementary assets of existing producers. Calvano (2007) argues that: "we highlight the role of destruction rather than creation in driving innovative activity. The formal analysis shows that destructive creation unambiguously leads to higher profits whatever the innovation cost". In particular, disruptive innovations disturb the business models of incumbents that have to counter mobilize resources to sustain their competitive advantage in the presence of market change (Garud et al., 2002; Markman and Waldron, 2014). In fact, new radical technologies in markets require that incumbents undertake specific R&D investments and strategic change to support competitive advantage (Christensen and Raynor, 2003; cf., Gioia and Chittipeddi, 1991; Teece et al., 1997). Current R&D management of incumbents, to support innovation processes, is more and more based on network organizations to build research alliances and strategic partnerships for increasing the access to external knowledge from new firms and/or research organizations (cf., Coccia, 2016b; Nicholls-Nixon and Woo, 2003). Kapoor and Klueter (2015) argue that incumbents tend to not invest in disruptive technological regimes and maintain a competence-enhancing approach. In some industries, such as biopharmaceutical sector, current wave of research alliances and acquisitions may help incumbents to overcome this "inertia" both in the initial stage of research and in the later stage of development. Other studies show that R&D investments of innovative enterprises in pharmaceutical industry are directed towards both internal research units and strategic alliances to accelerate the drug discovery process (Coccia, 2014).

However, theoretical framework of disruptive technologies suffers of some limitations, such as the ambiguity in the definition of disruptive innovations that considers technologies but also products and business models (*cf.,* Christensen and Raynor, 2003; Tellis, 2006). Strictly speaking, a disruptive technological innovation is fundamentally a different phenomenon from a disruptive business-model innovation. Disruptive innovations arise in different ways, have different competitive effects, and require different responses into the organizational behaviour of incumbents and entrants (Markides, 2006, p. 19). This diversity can be due to a variation in the sources of innovation, such as in some industries, users develop innovation, in other sectors, innovations are due to suppliers of related components and product manufactures (von Hippel, 1988). A vital factor in the development of innovations is also played by the coevolution of technical and institutional events (Van de Ven and Garud, 1994). The theory of disruptive technologies also seems to show some inconsistencies in many markets because new small entrants can generate new technology and innovations but their development and diffusion in markets present many economic barriers, such as within biopharmaceutical industry (Coccia, 2014; 2016; cf. also Calabrese et al., 2005; Cavallo et al., 2014). In short, the theory of disruptive technologies presents some difficulties to explain the general drivers of technological and economic change.

This study here suggests the vital role of specific firms, called *disruptive firms* that in the ecosystems can generate and spread new technologies with market shifts within and between industries. The study proposes some characteristics of these disruptive firms that can clarify, as far as possible, a main source of innovation to explain drivers of technological change and, as a consequence, industrial, economic and social change.

The model of this study is in Figure 1. Unlike theoretical framework of disruptive innovation (Christensen, 1997), the theoretical framework here suggests that, leading firms -called disruptive firms-support the emergence and diffusion of new technology and radical innovations that generate market shifts, technological and economic change.

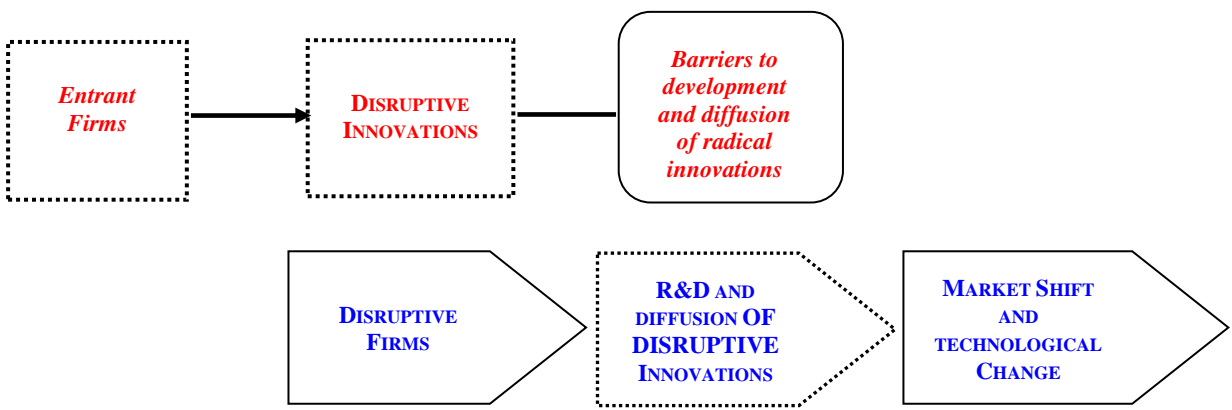

**Figure 1:** Disruptive firms sustain technological and economic change with the introduction and diffusion of technical breakthroughs.

The purpose of the present study is to see whether case study research supports the hypothesis that one of the general sources of technological change is due to disruptive firms (subjects) that generate market shifts, rather than disruptive technologies (objects) *per se*.

**Methods: case study research**

The methodology is based on an inductive analysis of case study research (Eisenhardt, 1989; Eisenhardt and Graebner, 2007).

The study analyzes the managerial and organizational behavior of specific leading enterprises (disruptive firms) to explain one of the general sources of technological and economic change. The firms under study are:

- Apple Inc. for Information and Communication Technologies (ICTs)

- AstraZeneca for biopharmaceutical industry

In particular, the hypothesis of this study is that specific and distinct firms, called disruptive firms, are the driving force of market shift in industries by introducing new products, standard and/or components in markets with new technology and innovation, generating technological and socioeconomic change (cf., Coccia 2009, 2012a, 2015a; Coccia and Wang, 2016). Of course, the emergence of a disruptive technology is a necessary but not sufficient condition for the development and diffusion of new technology in markets that generate industrial change. Manifold factors also create important conditions for supporting technical breakthroughs. This study here focuses on specific subjects, the disruptive firms that play a vital role in competitive markets. In order to support the theoretical framework, firstly, the study analyzes shortly these firms and then we contextualized the theory with some examples of new technology and the organizational and managerial behavior of disruptive firms that generate market shift, technological and economic change.

**Inductive analysis**

- Apple Inc. is an American multinational technology company headquartered in California (USA) that designs, develops, and sells consumer electronics, computer software, and online services. Apple was founded in 1976 to develop and sell personal computers. It was incorporated as Apple Computer Inc. in 1977, and was renamed as Apple Inc. in 2007 to reflect its shifted focus toward consumer electronics (Wozniak, 2007). Number of employees as of October 2016 is about 116,000 units.

    *Apple Inc. is a disruptive firm of storage devices.* A simple storage device was the floppy disk: a disk storage medium composed of a disk of thin and flexible magnetic storage medium encased in a rectangular plastic carrier. In 1983 Sony introduced 90 mm micro diskettes (better known as 3.5-inch -89 mm- floppy disks), which it had developed at a time when there were 4" floppy disks, and a lot of variations from different companies, to replace on-going 5.25" floppy disks. Apple Computer, a market leader in ICTs, decided to use in 1984 the 3½-inch drives produced by Sony in the Macintosh 128K model. This firm strategy effectively makes the 3½-inch drive a *de-facto* standard in markets. This Apples' decision generated a main market shift

and the format 3.5" floppy disks became dominant. Floppy disks 3.5" remained a popular medium for nearly 40 years, but their use was declining by the mid-1990s (Mee and Daniel, 1996). In 1998, Apple Inc. released the iMac G3 with a new store device, called USB because it considered the floppy disk an old technology. USB—or Universal Serial Bus—is a protocol for connecting peripherals to a computer. The development of the first USB technology began in 1994 by Intel and the USB-IF (USB Implementers Forum, Inc., formed with industry leaders like Intel, Microsoft, Compaq, LSI, Apple and Hewlett-Packard). USB was designed to standardize the connection of computer peripherals (Cunningham, 2014). The USB 1.0 debuted in late 1995 and transferred data at a rate of 12 megabits per second. This parasitic technology is associated to other host technologies, such as PCs. Interaction between these high-tech devices and a host computer without the need to disconnect or restart the computer also enables USB technology to render more efficient operation. As just mentioned, in 1998, the iMac G3 was the first consumer computer to discontinue legacy ports (serial and parallel) in favor of USB. This implementation helped to pave the way for a market of solely USB peripherals rather than those using other ports for devices. The combination of the ease of use, self-powering capabilities and technical specifications offered by USB technology and related devices helped this new technology to triumph over other port options (Au Yong, 2006; Tham, 2011). This decision of Apple generated a market shift and industrial change. In the presence of this technological change generated by a market leader, the ICT industry's reaction is to follow Apple's technological pathway, such as Dell, Hewlett-Packard, etc. that dumped the floppy drivers from their standard PCs. Trek Technology and IBM began selling the first USB flash drives commercially in 2000. IBM's USB flash drive had a storage capacity of 8 MB, more than five times the capacity of the then-common 3½-inch floppy disks (of 1440 KB). Similar pathway is with the Compact Disc (CD), a digital optical disc data storage format released in 1982 and co-developed by Philips and Sony (BBC News, 2007). The format was originally developed to store and play only sound recordings but was later adapted for storage of data (CD-ROM). Apple Inc. released the third generation of MacBook Pro in 2012 with a 15-inch screen that was a quarter thinner than its predecessor and the Retina Display with a much higher screen resolution. The MacBook Pro with Retina Display does not have an optical drive and to play discs, it is necessary to have an external Super Drive. This decision of a market leader generated a further market shift and industrial change towards new storage devices with the USB port, micro-USB or USBType-C (Hruska, 2015; Mee and Daniel, 1996; Goda and Kitsuregawa, 2012, USB, 2005).

*Apple Inc. is also a disruptive firm of wired headphones*. Headphones are pair of small listening devices that are electroacoustic transducers, which convert an electrical signal to a corresponding sound in the user's ear. They are designed to allow a single user to listen to an audio source privately. Firstly, the headphone with jack was created in the period 1890-1910 and with several generations is still used in many electronic devices. The study here focuses on a critical period associated to Bluetooth technology (a wireless technology standard for exchanging data over short distances from fixed and mobile devices, and building personal area networks-PANs). In fact, the revolution of ICT has generated several innovations such as the Bluetooth technology in 1999 (Bluetooth, 2017). The evolution of this technology has generated in 2004 the Bluetooth 2.0 with an Enhanced Data Rate for faster data transfer, in 2010 Bluetooth 4.0 with low energy and so on (Bluetooth, 2017). The interaction between Bluetooth and mobile phone has generated in 2002 the first mobile phone with integrated Bluetooth by Nokia, whereas the interaction between Bluetooth and headphones has also generated in 2003 the first Nokia headset, which was sold to end-users (Windows, 2012). The 29 June, 2007 Apple Inc. launched the 1st generation of *i*Phone with Bluetooth 2.0; the diffusion of the *i*Phone worldwide plays a main role in the evolution of several ICTs, driven by Apple Inc., which is one of the market leaders in smartphones and other mobile devices. In 2011, Apple Inc. has announced that new *i*Phone 4S supports Bluetooth 4.0 with low energy phone. In September 2016, the *i*Phone 7 of Generation 10$^{th}$ is launched without headphone jack 3.5mm. This strategic decision by Apple Inc. has a main impact for the evolution of new generations of headphones that will be more and more wireless to function, interact and survive with mobile devices (Coccia, 2017a). This decision of Apple Inc. to produce a new

*i*Phone 7 without jack 3.5mm for headphone generates a selection pressure on manufacturers of these technologies that are focusing on new technological directions of headphones with Bluetooth™ technology (wireless) generating an on-going technological substitution and "Destructive creation" (Calvano, 2007) of current headphones with wire. In short, this case study *seems* to confirm that new technologies and technological trajectories are driven by specific firms that play a role of destruction of current technologies in favor of the creation of new technology and technological standards. Other examples of the organizational behavior of Apple Inc. as disruptive firm, are the destruction of the physical keyboard in smartphones with the creation of virtual keyboards in the iPhone of 1$^{st}$ generation in 2007. In general, disruptive firms have the market power to support new technological trajectories and industrial change. In short, the innovative behavior of market leaders can be a main driving force of technological, industrial and economic change. Moreover, market shifts are due to leader firms of host technologies, such as PC or smartphones, rather than leader firms of parasitic technologies, such as headphones, storage devices, etc. (cf., Coccia, 2017a).

- AstraZeneca (AZ) is a British–Swedish research-based biopharmaceutical company. It is originated by a merger in 1999 of the Astra AB company formed in 1913 (Sweden) and British Zeneca Group formed in 1993. AstraZeneca (AZ) is a large corporation that has a net income of US$3.406 billion (AstraZeneca, 2016), total assets for US$60.12 billion (Forbes, 2016) and total number of employees for about 50,000 (AstraZeneca, 2015). The human and economic resources invested in R&D by AstraZeneca are about 15,000 units of personnel and over US$4 billion in eight countries (AstraZeneca, 2015). One of the research fields of AZ is anticancer treatments, such as for lung cancer. The current therapeutic treatments (technology) for advanced non-small cell lung cancer (NSCLC) are again mainly based on chemotherapy agents. However, this technology has low efficacy for lung cancer treatment since the mortality rate is still high (Coccia, 2014). AstraZeneca as incumbent firm in drug discovery industry has generated a main radical innovation to treat lung cancer: the target therapy Iressa® that is based on the blocking agent Gefitinib. These path-breaking anticancer drugs are generating a revolution in therapeutic treatments of lung cancer with mutation

Epidermal Growth Factor Receptor (EGFR) because they block specific enzymes and growth factor receptors involved in cancer cell proliferation (Coccia, 2012, 2014, 2016). Studies in the biology show that lung cancer can become resistant to these new drugs because of a secondary mutation (T790M) that generates a progression of the cancer with several metastases and, as a consequence, high mortality within five years (Coccia, 2012). Clovis Oncology is a small pharmaceutical company, which is generating innovative products for new treatments in oncology. Clovis was founded in 2009 and is headquartered in Boulder, Colorado. This small pharmaceutical firm, Clovis oncology, has generated a new technology to treat lung cancer with mutation T790M: a new target therapy for EGFR-mutant lung cancer (Clovis Oncology, 2015). However, this small firm has difficulties in the development of this radical innovation in a sector with high capital intensity for R&D. This problem has induced Clovis oncology to enter in the stock exchange to gather financial resources directed to support R&D of several innovative products in its pipeline. The structure of the sector based on larger corporation has induced the biopharmaceutical company AstraZeneca (2015) to introduce a similar innovation for mutant lung cancers, called Tagrisso™ (AZD9291), that it was approved by US Food and Drug Administration in 2015 (AstraZeneca, 2016). This case study also confirms the vital role of large and leader firms, in competitive markets based on high intensity of R&D, that have the power to generate and/or to spread path-breaking innovations in order to achieve and sustain competitive advantage, as well as the goal of a (temporary) profit monopoly to support their market shares and industrial leadership.

Next section endeavors to detect the general characteristics of these disruptive firms that generate technological, industrial and economic change.

**Discussion**

A main goal of this study is the concept of disruptive firms: they are firms with market leadership that deliberate introduce new and improved generations of durable goods that destroy, directly or indirectly, similar products present in markets in order to support their competitive advantage and/or market leadership (cf., Calvano, 2017). These disruptive firms support technological and industrial change and induce consumers to repeat their purchase in order to adapt to new socioeconomic environment. Firm strategy of these leading firms is directed to support innovation and market leadership with new technology. An example of disruptive firms is Apple Inc. that has the following organizational behaviour (cf., Backer, 2013; Barney, 1986; Fogliasso and Williams, 2014; Heracleous, 2013; O'Reilly et al., 1991; Schein, 2010).

1. A main and central leader in the organization, represented in the past by the founder Steve Jobs and subsequently by the CEO Tim Cook (Apple Inc., 2017). The hierarchy in Apple's organizational structure supports strong control over the organization that empowers top leader to control everything in the organization. This organizational behavior generates limited flexibility of lower levels of the hierarchy to respond to custom needs and market demand but it provides a clear leadership for R&D and strategic management of innovative products.

2. A large market share in mobile technology and associated industrial leadership. Samsung is the largest vendor in smartphones but it only captured 14% of smartphone profits, while Apple Inc. gathered 91% of them in 2015. Apple holds nearly 45% of the U.S. OEM (Original Equipment Manufacturer) market, and in a distant second is Samsung Electronics with 28% of the market. Notably, Apple is one of the only companies to actually advance its market share (from October through January), from 42.3% to 44.6%, for a 2.3% gain. Samsung's market share declined 2% from 30% in late 2016. Apple's *i*Phone accounted for 34% of all smartphone activations in the U.S. last quarter, leading all other smartphone brands. Samsung was just behind the *i*Phone at 33%, followed by LG at 14% share of activations (Kilhefner, 2017).

3. Founded in 1976, more than 40 years ago. The firm has a long presence and experience in the sector of computer hardware, software and electronics.

4. Headquarters is localized in a high-tech region, California, of a powerful country with socioeconomic influence on wide geoeconomic areas.

5. Apple's organizational culture is also highly innovative to support firm's product development processes and firm's industry leadership. Creativity and excellence are especially important in Apple's rapid innovation processes. Moreover, secrecy is part of the company's strategy to minimize theft of proprietary information or intellectual property. Apple employees agree to this organizational culture of secrecy, which is reflected in the firm's policies, rules and employment contracts. This aspect of Apple's organizational culture helps protect the business from corporate espionage and the negative effects of employee poaching. These characteristics of the company's organizational culture are key factors that enable success and competitive advantage (cf. also, Csaszar, 2013; Damanpour and Aravind, 2012, Lehman and Haslam, 2013).

Some characteristics of the organizational behavior of AstraZeneca (AZ) are (Coccia, 2014a, 2015, 2016a):

1. A characteristic similar to previous firm is a long experience in the market and leadership position in specific segments of the biopharmaceutical sector. In fact, Astra AB formed in 1913 (Sweden) and British Zeneca Group formed in 1993. Moreover, AstraZeneca is a large corporation in industry.

2. Higher specialization of technological capability in new research fields of genetics, genomics and proteomics to support drug discovery process.

3. Another characteristic of AZ is a division of scientific labour (cf. 'division of innovative labour' by Arora and Gambardella 1995; Coccia, 2014a). R&D strategy of this incumbent firm is to create strategic alliances with emerging firms for a division of scientific labour directed to reinforce and accelerate discovery process. In fact, AZ has strategic partnerships with organizations to complement in-house technological and scientific capabilities. In this manner, AZ supports rational modes of drug discoveries by integrative capabilities developed in collaboration with biotechnology firms (cf., Coccia, 2016b; Henderson 1994, pp. 607ff; Paruchuri and Eisenman, 2012). In particular, AZ builds and reinforces the scientific capabilities by strategic alliances with external sources of innovation: i.e., partnership with academic institutions, biotechs and other pharmaceutical companies to share skills, knowledge and resources through all phases of R&D process. In addition, the acquisition of the biotechnology firm MedImmune has improved and enlarged the R&D function and technological capabilities (AstraZeneca, 2015). This R&D management of AZ organizes the R&D labs with a network structure based on strategic alliances for supporting the process of disruptive innovations (figure 2). Network R&D organization reinforces the integrative capabilities in scientific fields, collective and cumulative learning between in-house R&D and external sources of innovation. Moreover, network structure of R&D generates a multiplicity of scientific stimuli and the adoption of different and complementary R&D management approaches (cf., Coccia, 2014a, 2016b; Henderson, 1994; Jenkins, 2010).

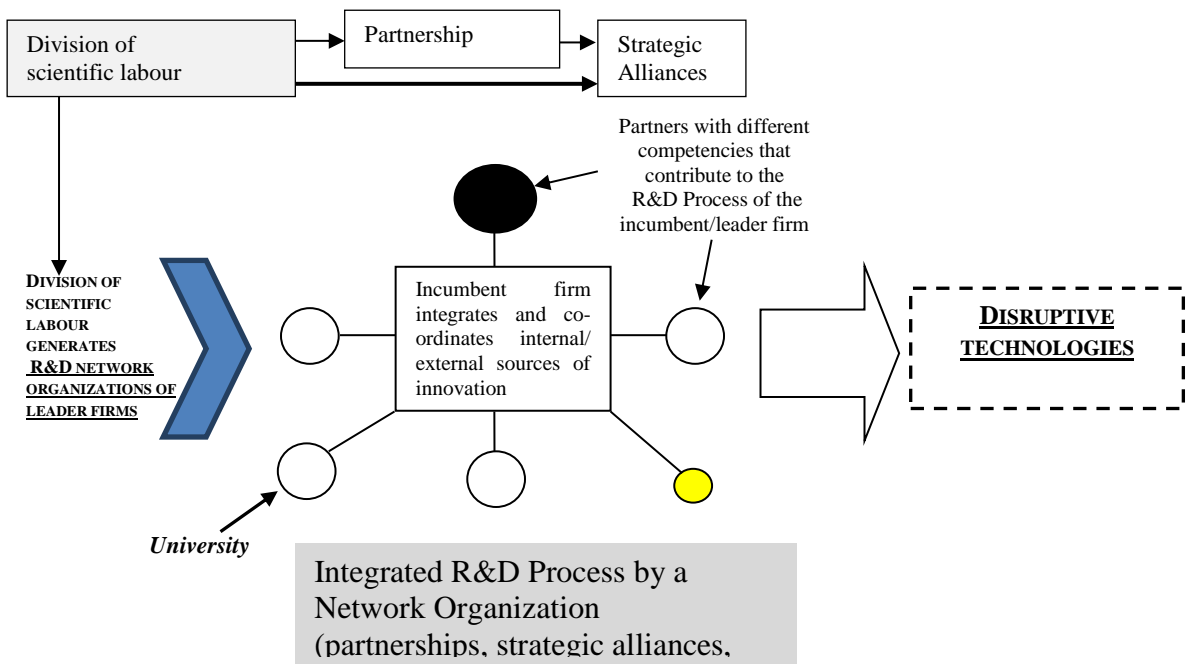

**Figure 2:** Network of R&D function of disruptive firms to support new technologies in innovative industries.

**Generalization of characteristics of disruptive firms that generate technological and industrial change**

The industrial dynamics shows that the theory of disruptive technology seems to be not consistent for explaining the R&D and diffusion of major innovations in main sectors such as ICTs and biopharmaceuticals. The inductive study here suggests that patterns of technological innovations in markets are dominated by incumbents rather than entrant firms, which have not the market power and structure to support path-breaking innovations across markets (Coccia, 2014a, 2015, 2016b, Daidj, 2016; Liao, 2011). In short, this study proposes the shift of the locus of one of basic causes of technological change, from disruptive technologies to disruptive firms that support path-breaking innovations and market shifts.

The case study research here reveals some general characteristics of disruptive firms that generate technological change. In particular,

- Large size, associated to a strong market power that supports an industrial leadership.

- Disruptive firms can or cannot generate radical and/or incremental innovations but they have the market power to spread and support new technology in markets generating industrial change.

- Forward-looking executives seeking to pioneer radical innovations in competitive markets.

- High R&D investments to lead the markets towards new technological trajectories, sustain competitive advantage, the goal of a (temporary) profit monopoly and industrial leadership.

- A long historical presence and expertise in the industry for many years (e.g., more than 40 years). The historical development path in industries supports the accumulation of technological knowledge, technical expertise and experience in the sector, more and more important for R&D and strategic management.

- Organizational and managerial behavior based on competence-destroying and competence-enhancing.

- Strong dynamic capabilities based on combinations of competences and resources that can be developed, deployed, and protected in order to stress exploiting existing internal and external firm specific competences and to address changing environments.

- R&D organization of disruptive firms is more and more based on a division of scientific labour. Network R&D organizations reinforce integrative capabilities, collective and cumulative learning between in-house R&D and external sources of innovation. Moreover, strategic alliances and partnership with innovative firms, university labs and suppliers support learning processes, accumulation of new knowledge and acceleration of innovation processes.

**Concluding Observations**

The theoretical framework of disruptive technologies seems that does not explain the dynamics of technological and economic change (*cf.,* Christensen, 1997). The study here endeavors to clarify, whenever possible, one of driving forces of technological change based on the role of leader firms, called *disruptive firms*. The central contribution of this work is an approach that integrates current frameworks in management and industrial organization to explain the sources of industrial and technological change (Cooper 1990; Dosi, 1988; O'Reilly III and Tushman, 2004; 2008).

In general, firms have goals, such as achieve and sustain competitive advantage (Teece et al., 1997).

One of the main organizational drivers of disruptive firms is the incentive to find and/or to introduce innovative solutions in new products, using new technology, in order to reduce costs, achieve and support the goal of a (temporary) profit monopoly and market (industrial) leadership. Case study research here also shows that R&D management of leading firms has more and more a division of scientific labour directed to accelerate innovation process and develop new technology. Disruptive firms generate significant shifts in markets with an ambidexterity strategy based on competence-destroying and competence-enhancing (cf., Danneels, 2006; Henderson, 2006; Hill and Rothaermel, 2003; Tushman and Anderson, 1986). Moreover, a main role in disruptive firms is also played by "forward-looking executives seeking to pioneer radical or disruptive innovations while pursuing incremental gains" (O'Reilly III and Tushman, 2004, p. 76). In general, disruptive firms, generating path-breaking innovations, grow more rapidly than other ones (Tushman and Anderson, 1986, p. 439).

On the basis of the argument presented in this paper, based on a case study research, we can therefore conclude that one of principal sources of technological and economic change is due to leading subjects, disruptive firms, which can be the distal sources of disruptive innovations in competitive markets, *ceteris paribus*. Disruptive firms have specific dynamic capabilities that generate learning processes, a vital cumulative change and path dependence in innovative industries (cf., Garud et al., 2010; Teece *et al.,* 1997).

The results of the analysis here are that:

(1) The conceptual framework here assigns a central role to leading firms (subjects) –disruptive firms- rather than disruptive technologies (objects) to sustain technological and economic change.

(2) *Disruptive firms* are firms with market leadership that deliberate introduce new and improved generations of durable goods that destroy, directly or indirectly, similar products present in markets in order to support their competitive advantage and/or market leadership. These disruptive firms support technological and industrial change and induce consumers to buy new products to adapt to new socioeconomic environment.

(3) The establishment and diffusion of disruptive technologies in markets are mainly driven by incumbent (large) firms with a strong market power. However, small (entrant) firms can generate radical innovations but they have to cope with high economic resources needed for developing new technology (*cf.,* Caner *et al.,* 2016). This financial issue explains the strategic alliances and partnerships between some incumbent and entrant firms to develop disruptive technologies. These collaborations mark a new phase in business development of innovations.

(4) Finally, the conceptual framework here also shows that R&D management of disruptive firms is more and more based on a division of scientific labor directed to reinforcing the integrative capabilities and collective learning between internal and external sources of innovation in order to accelerate discovery process.

Overall, then, the conceptual framework here, has several components of generalization that could easily be extended to explain the source of technological and economic change. To conclude, this study suggests that one of principal sources of industrial change is due to disruptive firms in competitive markets. To put it differently, this study provides a preliminary analysis of driving forces of technological change based on disruptive firms rather than disruptive technologies *per se*. However, the conclusions of this study are of course tentative. Most of the focus here is based on a case study research, clearly important but not sufficient for broader understanding of the complex and manifold sources of technological change. Moreover, the evidentiary basis of this paper is also weak, but this study may form a ground work for development of more sophisticated theoretical and empirical analyses to explain, whenever possible general causes of the technological and economic change. Hence, there is need for much more detailed research to explain the reasons for technological change in industries because we know that, in competitive markets with market dynamism, other things are often not equal over time and space. In fact, Wright (1997, p. 1562) properly claims: "In the world of technological change, bounded rationality is the rule".